\documentstyle[preprint,aps,epsf]{revtex}
\def\be{\begin{equation}}
\def\ee{\end{equation}}
\def\bc{{\bf c}}
\def\bh{{\bf h}}
\def\bt{{\bf t}}
\def\bE{{\bf E}}
\def\bM{{\bf M}}
\def\bEh{\tilde {\bf E}}
\def\bzh{\tilde {\bf z}}
\def\bah{\tilde {\bf a}}

\def\bAh{\tilde {\bf A}}
\def\bQ{\hat {\bf Q}}
\def\bQr{{\bf Q}}
\def\bq{{\bf q}}
\def\bS{{\bf S}}
\def\bJ{{\bf J}}

\def\bXh{{\bf {\tilde X}}}

\def\bg{{\bf g}}

\def\rob{\mbox {\boldmath ${\rho}$}}

\def\balpha{\mbox {\boldmath ${\alpha}$}}
\def\bbeta{\mbox {\boldmath ${\beta}$}}

\def\bgamma{\mbox {\boldmath ${\gamma}$}}
\def\chat{{{\bf {\hat c}}}}
\def\hhat{{{\bf {\hat h}}}}

\def\nablab{\mbox {\boldmath ${\nabla}$}}

\begin{document}
\voffset -0.1in
\title{Multicomponent mixture of charged hard-sphere chain molecules\\
in the polymer mean-spherical approximation}
\author{Yu. V. Kalyuzhnyi}
\address{Institute for Condensed Matter Physics, Svientsitskoho 1, 290011
Lviv, Ukraine}
\author{P. T. Cummings}
\address{Department of Chemical Engineering, University of Tennessee, 
Knoxville 37996-2200, USA}

\maketitle
\vspace{90mm}

{\it Running title}: Charged hard-sphere chain fluid 

{\it The address for correspondence}: Institute for Condensed Matter 
Physics, \\ Svientsitski Str. 1, 79011 Lviv, UKRAINE; e-mail:
yukal@icmp.lviv.ua

\clearpage

\begin{abstract}
\tighten

The analytical solution of the recently proposed ideal chain polymer 
mean-spherical approximation (Yu.Kalyuzhnyi, {\it Mol.Phys.}, {\bf 94}, 
735(1998)) is presented for the multicomponent mixture of charged 
hard-sphere linear chain flexible molecules. The solution apply to any 
mixture of chain molecules with arbitrary distribution of the charge and 
size of the beads along the molecular backbone. Closed form analytical 
expressions for the internal energy, Helmholtz free energy, chemical 
potentials and pressure are derived. By way of illustration thermodynamical 
properties of several versions of the fluid of charged chain molecules of 
different length, including the molecules with uniform, diblock and 
alternating distribution of the charge are studied. Theoretical predictions 
are in reasonable agreement with available computer simulation predictions. 
We present also the liquid-gas phase diagrams for the systems with diblock 
and alternating distribution of the charge.

\end{abstract}
\narrowtext

\vfill\eject 
\section{Introduction} 

\label{intro}

In recent years substantial amout of efforts has been focused on the 
development of theoretical approaches predicting the equilibrium 
structure and thermodynamical properties of charged polymer solutions. This 
information enables one to esteblish the stability limits and to build the 
phase diagrams of such fundamentally and technologically importent systems 
as polyelectrolyte solutions (for example DNA, polyacrilic acid) and 
polyampholyte solutions (for example proteins). One of the simplest 
continuum models of charged polymer solutions is represented by the fluid of 
charged hard-sphere linear chain flexible molecules. In spite of its 
simplicity this model incorporates several essential features of charged 
polymer fluids, such as excluded volume effects, chain connectivity and 
flexibility, distribution of the charge along the chain backbone and Coulomb 
interaction. Recently several off-lattice theories for the fluid of flexible 
charged chain molecules, based on the integral-equation techniques developed 
for the fluids of small molecules, have been proposed. These include 
extension \cite{belloni} of the polymer reference interaction site model 
(RISM) integral equation theory of Schweizer and Curro 
\cite{currev}, extension \cite{china1,china2} of the simple interpolation 
scheme (SIS) of Stell and Zhou \cite{stellzhou} and 
extensions  \cite{kstellion,kblum0,kblum,ktd,solms} of the multidensity 
integral equation theory for associating fluids 
of Wertheim \cite{w34}. The latter 
studies are based on the analytical solution of the mean spherical 
approximation (MSA) version of the multidensity theory, the so-called 
polymer MSA (PMSA) \cite{kstellion,kblum0,kblum,holkal}. In the case of 
uncharged system direct application of Wertheim's original formulation by 
Chang and Sandler \cite{chang1,chang2} gave 
rise to a polydisperse fluid of chains with a prescribe mean number of 
beads. A subsequent extension by Kalyuzhnyi and Cummings \cite{kalcum} 
yields a model that polymerizes to a fluid of chains of fixed length in the 
complete association limit. In that limit, two-density Wertheim's 
Ornstein-Zernike (OZ) equation becomes identical to the Rossky-Chiles 
\cite{rossky} version of the "proper" site-site integral equations 
\cite{csl} ("proper-RISM") as first noted by Kalyuzhnyi and Stell 
\cite{kstellsym} and later discussed in more details by Stell 
\cite{stellphys} and by Kalyuzhnyi and Cummings \cite{kalcumon}. More 
recently complete association version of this extension was combined with 
the ideal chain approximation \cite{w34,chang1,kallin1} and generalized for 
the systems with arbitrary long-range potential outside the hard-core region 
\cite{ktd}. As a result the corresponding version of the PMSA was formulated 
and H$\o$ye-Stell scheme \cite{hoyestell} of calculating the simple fluids 
thermodynamical properties was extended in the frames of the PMSA 
\cite{ktd}. Subsequently von Solms and Chiew \cite{solms} combine PMSA and 
ideal chain approximation in a way similar to that proposed by Chang and 
Sandler \cite{chang1}. The resulting polydisperse chain mixture was used to 
describe thermodynamical properties of the two-component mixture of 
fixed-length chain molecules with equally charged beads and oppositelly 
charged monomer counterions. Results of the theory appeare to be in 
reasonable agreement with computer simulation results \cite{kremer}. We note 
in passing, that ideal chain approximation utilized by von Solms and Chiew 
\cite{solms} is somewhat different from the original one \cite{w34,chang1}. 
In the corresponding Wertheim's OZ equation the authors take into account 
all partial correlation functions, while due the original ideal chain 
approximation \cite{w34,chang1} correlations involving at least one doubly 
bonded particle are neglected. 
More recently general solution of the PMSA \cite{kblum0,kblum} for 
polymerizing charged hard spheres, supplemented by the ideal chain 
approximation, was used to derive closed form analytical expressions for 
thermodynamical properties of charged chain fluid with a certain 
restrictions imposed on the sizes of the chain beads \cite{berblumpes}. In 
the latter study the authors propose an improved version of the ideal chain 
approximation, which satisfies the Debye-H\"uckel limiting law.

In this paper we present an analytical solution of the ideal chain PMSA 
\cite{ktd} for the multicomponent mixture of charged hard-sphere flexible 
chain molecules. Unlike the previous studies proposed version of 
the PMSA is quite general and apply for any mixture of the chain molecules 
with arbitrary distribution of the charge and size of the beads along the 
molecular backbone. Our solution represents a complete association limit of 
the general solution obtained earlier \cite{kblum0,kblum} for the fluid of 
polymerizing charged hard spheres. Similar, as in the case of primitive 
electrolyte model \cite{blumrpm} this solution reduces to the solution of 
only one algebraic equation for the MSA-like screening parameter $\Gamma$. 
We also generalize recently developed method of calculating PMSA 
thermodynamics via the energy route \cite{ktd} for the multicomponent case 
and derived closed form analytical expressions for internal energy, 
Helmholtz free energy, chemical potentials and pressure in terms of the 
present solution. By way of illustration we consider several different 
versions of the fluid of charged hard-sphere chain molecules, including the 
molecules with uniform, diblock and alternating distribution of the charge 
along the chain backbone.

The paper is orgenized as follows. In the next section we discuss detailes 
of the model to be studied and formulate ideal chain PMSA theory. Section 
III gives solution of the ideal chain PMSA and Section IV contains 
expressions for thermodynamic and structure properties of the system. 
Numerical results and discussion can be found in Section V and in Section VI 
we collect our conclusions.

\section{The model and ideal chain PMSA theory}

The model fluid we consider consists of the $M$-component mixture of freely 
jointed tangent hard-sphere chain molecules with each molecule of species 
$a$ represented by $m_a$ charged hard-sphere sites of sizes 
$\sigma_\alpha^a$ and charges $ez_\alpha^a$.
The total number density of the system is $\rho=\sum_a\rho_a$, where 
$\rho_a$ is the number density of species $a$, and we assume charged 
neutrality conditions
\be
\sum_{a\alpha}\rho_a z_\alpha^a=0
\label{chneutr}
\ee
We denote the molecular species by the small letters $a,b,c,...$ taking the 
values $1,2,...,M$ and the site type in a given molecule $a$ by a small 
greek letters $\alpha,\beta,\gamma,...$, which take the values 
$1,2,...,m_a$. Thus the site has two indices, one denoting the species of 
the molecule and the other the site type in the molecule. For example the 
hard-sphere diameter of the site of type $\alpha$ belonging to the molecule 
of the species $a$ is denoted by $\sigma_\alpha^a$.

The site-site pair potential $U_{\alpha\beta}^{ab}(r)$ between the sites of 
the type $\alpha$ and $\beta$ belonging to the molecules of species $a$ and 
$b$ can be written in the form
\be
U_{\alpha\beta}^{ab}(r)=U_{\alpha\beta}^{(hs)ab}(r)+
U_{\alpha\beta}^{(C)ab}(r)
\label{pot}
\ee
where $U_{\alpha\beta}^{(hs)ab}(r)$ is the hard-sphere potential and 
$U_{\alpha\beta}^{(C)ab}(r)$ is the Coulomb potential
\be
U_{\alpha\beta}^{(C)ab}(r)={e^2z_\alpha^az_\beta^b\over \epsilon_0r}
\label{culpot}
\ee
with $\epsilon_0$ being the dielectric constant of the continuum.

In this study we are using the multidensity version of the mean spherical 
approximation (MSA) \cite{w34,holkal,kstellion,ktd} (or polymer MSA (PMSA)) 
supplemented by the so-called ideal chain approximation 
\cite{w34,chang1,kallin1,ktd}. Both PMSA and the ideal chain approximation 
have been discussed at length earlier and therefore we will omit any details 
and present here only the final expressions. The theory consists of the 
Ornstein-Zernike-like integral equation
\be
\hhat_{\alpha\beta}^{ab}(k)=\chat_{\alpha\beta}^{ab}(k)
+\sum_c\rho_c\sum_\gamma\chat_{\alpha\gamma}^{ac}(k)\balpha
\hhat_{\gamma\beta}^{cb}(k)
\label{oz}
\ee
and PMSA boundary conditions
\be
\left\{
\begin{array}{lllll}
\bc_{\alpha\beta}^{ab}(r)&=&-\bE\beta U_{\alpha\beta}^{(C)ab}(r)+
{\bt_{\alpha\beta}^{ab}\over 
2\pi\sigma_{\alpha\beta}}
\delta (r-\sigma_{\alpha\beta}^{ab}),\hspace{5mm}
&r&>\sigma_{\alpha\beta}^{ab}={1\over 2}
\left(\sigma_\alpha^a+\sigma_\beta^b\right)\\
\bh_{\alpha\beta}^{ab}(r)&=&-\bE, 
&r&<\sigma_{\alpha\beta}^{ab}
\end{array}
\right.
\label{amsa}
\ee
where  
$\hhat_{\alpha\beta}^{ab}(k)$ and $\chat_{\alpha\beta}^{ab}(k)$
are the matrices with the elements being the Fourier transforms of the elements of
the matrices $\bh_{\alpha\beta}^{ab}(r)$ and $\bc_{\alpha\beta}^{ab}(r)$
$$
\bh_{\alpha\beta}^{ab}(r),\bc_{\alpha\beta}^{ab}(r)=
\pmatrix{
c_{\alpha_0\beta_0}^{ab}(r)&c_{\alpha_0\beta_A}^{ab}(r)&
                                        c_{\alpha_0\beta_B}^{ab}(r) \cr 
c_{\alpha_A\beta_0}^{ab}(r)&c_{\alpha_A\beta_A}^{ab}(r)& 
                                        c_{\alpha_A\beta_B}^{ab}(r) \cr
c_{\alpha_B\beta_0}^{ab}(r)&c_{\alpha_B\beta_A}^{ab}(r)&
                                        c_{\alpha_B\beta_B}^{ab}(r) \cr},
$$
$\bt_{ij}$, $\balpha$ and $\bE$ 
are the following matrices
$$
t_{\alpha_i\beta_j}^{ab}={\delta_{ab}\over 2\rho_a}
\left[\delta_{iA}\delta_{jB}
{\delta_{\alpha\beta+1}\over \sigma_{\alpha\alpha-1}}+
\delta_{iB}\delta_{jA}
{\delta_{\alpha\beta-1}\over \sigma_{\alpha\alpha+1}}\right],
\hspace{5mm}
\alpha_{ij}=1-\delta_{ij}+\delta_{0i}\delta_{0j},
\hspace{5mm}
E_{ij}=\delta_{0i}\delta_{0j}.
$$
Here the lower indices $i,j$ each take the values $0,A,$ and $B$, and denote 
the bonding states of the corresponding particle, i.e. $0$ denotes 
the unbonded state and $A$ and $B$ denote the $A$-bonded and $B$-bonded 
states, respectively. The total partial correlation functions 
$h_{\alpha_i\beta_j}^{ab}(r)$ are related to the site-site total correlation 
functions $h_{\alpha\beta}^{ab}(r)$ by the following relation
\be
h_{\alpha\beta}^{ab}(r)=\sum_{ij}h_{\alpha_i\beta_j}(r)
\label{total}
\ee

The set of the OZ equations (\ref{oz}) together with the PMSA closure conditions
(\ref{amsa}) represent our ideal chain PMSA theory for the fluid of 
charged linear chain hard-sphere molecules.

\section{Solution of the ideal chain PMSA for the fluid of charged chain 
molecules}

Recently a general solution of the PMSA for the multicomponent mixture of 
polymerizing charged hard spheres was published \cite{kblum0,kblum}. Here 
we elaborate on this general solution by utilizing additionally the ideal 
chain approximation \cite{w34,chang1,ktd} and specializing it for the case of the model at 
hand.

The general solution was obtained using Baxter's tecnique, which factorizes 
the initial OZ equation (\ref{oz}) into a set of two equations
\be
\bS_{\alpha\beta}^{ab}(|r|)=\bQr_{\alpha\beta}^{ab}(r)-\sum_c\rho_c
\sum_\gamma\int 
dr'\bQr_{\alpha\gamma}^{ac}(r')\balpha 
\left[\bQr_{\beta\gamma}^{bc}(r'-r)\right]^T 
\label{ozf3}
\ee
\be
\bJ_{\alpha\beta}^{ab}(|r|)=\bQr_{\alpha\beta}^{ab}(r)+\sum_c\rho_c
\sum_\gamma\int 
dr'\bJ_{\alpha\gamma}^{ac}(|r'-r|)\balpha \bQr_{\gamma\beta}^{cb}(r')
\label{ozf4}
\ee
where $T$ denotes the transpose matrix. The projections 
$\bS_{\alpha\beta}^{ab}(r)$ and $\bJ_{\alpha\beta}^{ab}(r)$ 
\be
\bS_{\alpha\beta}^{ab}(r)=2\pi\int_r^\infty dr'r'{\bf 
c}_{\alpha\beta}^{ab}(r'), \hspace{5mm}
\bJ_{\alpha\beta}^{ab}(r)=2\pi\int_r^\infty dr'r'{\bf 
h}_{\alpha\beta}^{ab}(r') 
\label{proj}
\ee
satisfy the following boundary conditions
\be
\left\{
\begin{array}{lllll}
\bJ_{\alpha\beta}^{ab}(r)&=&\pi r^2\bE+\bJ_{\alpha\beta}^{ab},\hspace{10mm}
&r&\le\sigma_{\alpha\beta}^{ab}\\
\bS_{\alpha\beta}^{ab}(r)&=&-{\beta e^2\over \epsilon_0}{z_\alpha^az_\beta^b
e^{-\mu |r|}\over \mu}\bE,
&r&>\sigma_{\alpha\beta}^{ab}
\end{array}
\right.
\ee
obtained from (\ref{amsa}). Here 
$\bJ_{\alpha\beta}^{ab}=\bJ_{\alpha\beta}^{ab}(0)$ and the limit 
$\mu\rightarrow 0$ is to be taken at the end of the calculations.

From the analysis of the equations (\ref{ozf3}) (\ref{ozf4}) we get the 
following expression for the factor function $\bQr_{\alpha\beta}^{ab}(r)$
\be
\bQr_{\alpha\beta}^{ab}(r)=\theta\left(\sigma_{\alpha\beta}^{ab}-r\right)
\left[\bq_{\alpha\beta}^{ab}(r)+\bt_{\alpha\beta}^{ab}\right]-
\left(\bzh_\alpha^a\right)^T\bah_\beta^b,\hspace{7mm}
r>\lambda_{\alpha\beta}^{ab}
\label{qcap}
\ee
where 
$\lambda_{\alpha\beta}^{ab}=\left(\sigma_\alpha^a-\sigma_\beta^b\right)/2$ 
and the function $\bq_{\alpha\beta}^{ab}(r)$ is defined in the range
$\lambda_{\alpha\beta}^{ab}<r<\sigma_{\alpha\beta}^{ab}$ by
\be
\bq_{\alpha\beta}^{ab}(r)={1\over 
2}\bEh^T\bAh_\beta^b\left(r-\sigma_{\alpha\beta}^{ab}\right)
\left(r-\lambda_{\beta\alpha}^{ba}\right)+\beta_{\alpha\beta}^{ab}
\left(r-\sigma_{\alpha\beta}^{ab}\right).
\label{qsmall}
\ee
Here $\bEh,\bah_\alpha^a$ and $\bzh_\alpha^a$ are the row vectors
$$
\bEh=\left(1,0,0\right),\;\;\;
\bah_\alpha^a=\left(a_{\alpha_0}^a,a_{\alpha_A}^a,a_{\alpha_B}^a\right),\;\;\;
\bzh_\alpha^a=\left(z_\alpha^a,0,0\right).
$$

Coefficients of the Baxter $q$-function can be expressed in terms of only 
one parameter, i.e. the screening PMSA parameter $\Gamma$ \cite{blumrpm,kblum0,kblum}. We have
\be
\beta_{\alpha_i\beta_j}^{ab}=\delta_{i0}\delta_{j0}{\pi\sigma_\beta^b\over 
\Delta}+
{2\pi\beta^*\over \sigma_\alpha^a\Gamma}\left(X_{\alpha_0}^a-
\delta_{i0}z_\alpha^a\right)X_{\beta_j}^b,
\label{cof1}
\ee
\be
A_{\beta_j}^b={4\pi\beta^*\over \Gamma}\eta^BX_{\beta_j}^b+\delta_{j0}
{2\pi\over \Delta}\left[1+\zeta_2\sigma_\beta^b{\pi\over 2\Delta}\right]-
{\pi\over \Delta} \left[\left(1-\delta_{\beta m_a}\right)
{\delta_{jB}\sigma_{\beta+1}^b\over \sigma_{\beta+1\beta}^{bb}}+
\left(1-\delta_{\beta 1}\right)
{\delta_{jA}\sigma_{\beta-1}^b\over 
\sigma_{\beta-1\beta}^{bb}}\right]
\label{cof2}
\ee
\be
\Delta=1-{1\over 6}\pi\zeta_3,\hspace{4mm}
\zeta_n=\sum_d\rho_d\sum_k\left(\sigma_k^d\right)^n,\hspace{4mm}
\chi_2=\sum_d\rho_d\sum_kz_k^d\left(\sigma_k^d\right)^2
\label{cof3a}
\ee
\be
a_{\alpha_i}^a=2\pi\beta^*X_{\alpha_i}^a/\Gamma,
\label{cof3}
\ee
\be
X_{\alpha_0}^a=\left[z_\alpha^a-\eta^B\left(\sigma_\alpha^a\right)^2\right]
\Gamma_\alpha^a,\hspace{4mm}
X_{\alpha_i}^a=\sigma_\alpha^a\left[\tau_{\alpha_i}^a(z)-\eta^B
\tau_{\alpha_i}^a(\sigma^2)\right],\;\;\;(i\neq 0)
\label{cof45}
\ee
\be
\eta^B={{\pi\over 2\Delta}\sum_d\rho_d\sum_\gamma\sigma_\gamma^d\left\{
z_\gamma^d\Gamma_\gamma^d+\sigma_\gamma^d\left[\tau_{\gamma_A}^d(z)+
\tau_{\gamma_B}^d(z)\right]\right\}\over
1+{\pi\over 2\Delta}\sum_d\rho_d\sum_\gamma (\sigma_\gamma^d)^2\left[
\sigma_\gamma^d\Gamma_\gamma^d+\tau_{\gamma_A}^d(\sigma^2)+
\tau_{\gamma_B}^d(\sigma^2)\right]},
\hspace{6mm}
\Gamma_\alpha^a=\left(1+\sigma_\alpha^a\Gamma\right)^{-1}
\label{cof6}
\ee
\be
\tau_{\alpha_A}^a(y)={1\over 2}\sum_{\gamma=2}^\alpha\Gamma_{\gamma-1}^a
\rho_a^{\alpha-\gamma}{y_{\gamma-1}^a\over \sigma_{\gamma\gamma-1}^{aa}}
\prod_{\delta=\gamma}^\alpha\Gamma_\delta^a\left[2^{\gamma-\alpha}
\left(1-\delta_{\gamma\alpha}\right)\prod_{\omega=\gamma+1}^\alpha
{\sigma_{\omega-1}^a\over 
\sigma_{\omega\omega-1}^a}+\delta_{\gamma\alpha}\right]\left(1-\delta_{\alpha 
1}\right)
\label{cof7}
\ee
\be
\tau_{\alpha_B}^a(y)=
{1\over 2}\sum_{\gamma=\alpha}^{m_a-1}\Gamma_{\gamma+1}^a 
\rho_a^{\gamma-\alpha}{y_{\gamma+1}^a\over \sigma_{\gamma\gamma+1}^{aa}} 
\prod_{\delta=\alpha}^\gamma\Gamma_\delta^a\left[2^{\alpha-\gamma} 
\left(1-\delta_{\gamma\alpha}\right)\prod_{\omega=\alpha}^{\gamma-1} 
{\sigma_{\omega+1}^a\over 
\sigma_{\omega\omega+1}^a}+\delta_{\gamma\alpha}\right]\left(1-
\delta_{\alpha m_a}\right) 
\label{cof8}
\ee
Here $\Gamma$ satisfies the following equation
\be
\Gamma^2=\pi\beta^*\sum_{a\alpha}\rho_a\sum_\alpha
\bXh_\alpha^a\balpha\left(\bXh_\alpha^a\right)^T
\label{gamma}
\ee
where $\bXh_\alpha^a$ is the row vector 
$\bXh_\alpha^a=(X_{\alpha_0}^a,X_{\alpha_A},X_{\alpha_B})$.

Thus solution of the PMSA for the present model reduces to the solution of 
the algebraic equation (\ref{gamma}) for the screening parameter $\Gamma$.

\section{Structure and thermodynamics}

\subsection{Structure properties}

The contact values of the regular part of the partial pair distribution 
functions $g_{\alpha_i\beta_j}^{ab}(\sigma_{\alpha\beta}^{ab}+)$ follow from 
the relation (\ref{ozf4}) after differentiating it with respect to $r$ and 
taking the limit of $r\rightarrow \sigma_{\alpha\beta}^{ab}+$
\be
2\pi\sigma_{\alpha\beta}^{ab}
g_{\alpha_i\beta_j}^{ab}(\sigma_{\alpha\beta}^{ab}+)=
2\pi\delta_{0i}\delta_{j0}\sigma_{\alpha\beta}^{ab}
g_{\alpha\beta}^{(hs)ab)}(\sigma_{\alpha\beta}^{ab}+)-2\pi\beta^*
X_{\alpha_i}^aX_{\beta_j}^b+T_{\alpha_i\beta_j}^{ab}
\label{cont}
\ee
where $g_{\alpha\beta}^{(hs)ab}(\sigma_{\alpha\beta}^{ab}+)$ is the 
hard-sphere contact values and
$$
T_{\alpha_i\beta_j}^{ab}=\delta_{ab}\left\{\left[
{(1-\delta_{\alpha m_a})(1-\delta_{\alpha m_a-1})\delta_{iB}\delta_{jA}
\delta_{\alpha\beta-2}\over 4\rho_a\sigma_{\alpha\alpha+1}^{aa}
\sigma_{\alpha+1\beta}^{aa}}+
{(1-\delta_{\alpha 1})(1-\delta_{\alpha 2})\delta_{iA}\delta_{jB}
\delta_{\alpha\beta+2}\over 4\rho_a\sigma_{\alpha\alpha-1}^{aa}
\sigma_{\alpha-1\beta}^{aa}}\right]\right.+
$$
$$
{\pi\over 2\Delta}\delta_{j0}\sigma_\beta^a\left[(1-\delta_{\alpha 
m_a})\delta_{iB}{\sigma_{\alpha+1}^a\over \sigma_{\alpha\alpha+1}^{aa}}+
(1-\delta_{\alpha 1})\delta_{iA}{\sigma_{\alpha-1}^a\over 
\sigma_{\alpha\alpha-1}^{aa}}\right]+
$$
\be
\left.
{\pi\over 2\Delta}\delta_{i0}\sigma_\alpha^a\left[(1-\delta_{\beta 
m_a})\delta_{jB}{\sigma_{\beta+1}^a\over \sigma_{\beta+1\beta}^{aa}}+
(1-\delta_{\beta 1})\delta_{jA}{\sigma_{\beta-1}^a\over 
\sigma_{\beta-1\beta}^{aa}}\right]\right\}
\label{T}
\ee

The values of $g_{\alpha_i\beta_j}^{ab}(r)$ for 
$r>\sigma_{\alpha\beta}^{ab}$ can be calculated using the following relation
\be
g_{\alpha_i\beta_j}^{ab}(r)=\delta_{0i}\delta_{0j}+{1\over 2\pi^2r}
\int_0^\infty\left[{\hat 
\gamma}_{\alpha_i\beta_j}^{ab}(k)-\delta_{i0}\delta_{j0}\beta {\hat 
U}_{\alpha\beta}^{(C)ab}(k)\right]k\sin{(kr)}\;dk 
\label{gofr}
\ee
where ${\hat U}_{\alpha\beta}^{(C)ab}(k)$ is the Fourier transform of the 
Coulomb potential $U_{\alpha\beta}^{(C)ab}(r)$ and expression for the 
function ${\hat \gamma}_{\alpha_i\beta_j}^{ab}(k)=
{\hat h}_{\alpha_i\beta_j}^{ab}(k)-{\hat c}_{\alpha_i\beta_j}^{ab}(k)$ 
follows from the set of equations (\ref{ozf3}) and (\ref{ozf4}), written in 
the Fourier $k$-space
\be
\left\{
\begin{array}{lll}
\rob^{-1}-\chat(k)&=&\bQ(k)\rob\bQ^T(-k)\\
\rob^{-1}+\hhat(k)&=&\left[\rob\bQ(k)\rob\bQ^T(-k)\rob\right]^{-1}
\end{array}
\right.,
\label{ozk}
\ee
which gives
\be
\hat {\bgamma}(k)=\bQ(k)\rob\bQ^T(-k)+\left[\rob\bQ(k)\rob\bQ^T(-k)
\rob\right]^{-1}-2\rob^{-1}.
\label{hmc}
\ee
Here $\rob$, $\hhat(k)$, $\chat(k)$ and $bQ(k)$ are the matrices with the 
elements $\rho_{\alpha_i\beta_j}^{ab}=
\delta_{ab}\delta_{\alpha\beta}\alpha_{ij}\rho_a$, 
$\hat h_{\alpha_i\beta_j}^{ab}(k)$, $\hat c_{\alpha_i\beta_j}^{ab}(k)$ and
\be
{\hat Q}_{\alpha_i\beta_j}^{ab}(k)=
\left[\rob^{-1}\right]_{\alpha_i\beta_j}^{ab}-
\int_{\lambda_{\beta\alpha}}^\infty dr\;e^{ikr}Q_{\alpha_i\beta_j}^{ab}(r),
\label{qk}
\ee
respectively. Substituting expression for the fuction 
$Q_{\alpha_i\beta_j}^{ab}(r)$ (\ref{qcap}) into the right hand side of the 
Fourier transformation (\ref{qk}) we find that
$$
\bQ_{\alpha\beta}^{ab}(k)=\delta_{ab}\delta_{\alpha\beta}
\left(\rho_a\balpha\right)^{-1}-
\left\{\bEh\bAh_\beta^b\varphi_2(k,\sigma_\alpha^a)+\left(
\bbeta_{\alpha\beta}^{ab}-{1\over 2}\bEh\bAh_\beta^b\sigma_\alpha^a\right)
\varphi_1(k,\sigma_\alpha^a)+\right.
$$
\be
\left.
+\left(\bbeta_{\alpha\beta}^{ab}\sigma_\alpha^a-\bt_{\alpha\beta}^{ab}\right)
\varphi_0(k,\sigma_\alpha^a)-(\bzh_\alpha^a)^T\bah_\beta^b{i\over k}\right\}
\label{qkfin}
\ee
where 
$$
\varphi_2(k,\sigma)=-{1\over k^3}\left[\left({1\over 
2}k^2\sigma^2+ik\sigma-1\right)e^{ik\sigma}+1\right],\hspace{4mm}
\varphi_1={1\over k^2}\left[\left(1-ik\sigma\right)e^{ik\sigma}-1\right],
$$
\be
\varphi_0(k,\sigma)={i\over k}\left(e^{ik\sigma}-1\right)
\label{phi}
\ee
This result allow us to derive the expression for the structure factor
\be
{\hat S}_{\alpha\beta}^{ab}(k)=\delta_{ab}\delta_{\alpha\beta}+
\sqrt{\rho_a\rho_b}{\hat h}_{\alpha\beta}^{ab}(k)
\label{sfactor}
\ee
Using the second of the relations (\ref{ozk}) we have
\be
{\hat S}_{\alpha\beta}^{ab}(k)={1\over \sqrt{\rho_a\rho_b}}\left\{\left[
\bQ(k)\rob\bQ^T(-k)\right]^{-1}\right\}_{\alpha_0\beta_0}^{ab}
\label{sfacfin}
\ee

\subsection{Thermodynamic properties}

Expression for the excess internal energy $\Delta E$ follows from the 
standard relation
\be
{\Delta E\over V}=2\pi\sum_{ab}\rho_a\rho_b\sum_{\alpha\beta}\int_0^\infty
dr\;r^2U_{\alpha\beta}^{(C)ab}(r)g_{\alpha\beta}^{ab}(r),
\label{estandard}
\ee
which after some calculations gives
\be
\beta{\Delta E\over V}=\beta^*\sum_a\rho_a\sum_\alpha z_\alpha^a N_\alpha^a
\label{efin}
\ee
where $\sigma_\alpha^aN_\alpha^a=\sum_{i=0}^BX_{\alpha_i}^a-z_\alpha^a$

The other thermodynamical properties of the system in question can be 
calculated following the method developed earlier \cite{ktd}. This method 
generalizes H$\o$ye-Stell MSA energy route to thermodynamics \cite{hoyestell} in the 
frames of the present PMSA approach. Originally this generalization was 
developed for the one-component case. Extension of the method in the 
multicomponent case is rather straightforward and yields the following 
expressions for the Helmholtz free energy $A$, pressure $P$ and chemical 
potential $\mu_\alpha^a$ in excess to their reference system values
$$
-\beta {A-A^{(ref)}\over V}=J+{1\over 2}\sum_{ab}\rho_a\rho_b
\sum_{\alpha\beta}\left\{\balpha\left[\chat_{\alpha\beta}^{ab}-
\chat_{\alpha\beta}^{(0)ab}\right]\balpha\right\}_{00}
-\beta{\Delta E\over V}+
$$
$$
+{1\over 
3}\pi\sum_{ab}\rho_a\rho_b\sum_{\alpha\beta}
\left(\sigma_{\alpha\beta}^{ab}\right)^3 Tr 
\left[\bg_{\alpha\beta}^{ab}\balpha\bg_{\beta\alpha}^{ba}\balpha-
\bg_{\alpha\beta}^{(0)ab}\balpha\bg_{\beta\alpha}^{(0)ba}\balpha
\right]-
$$
\be
-{1\over 3}\sum_{ab}\rho_a\rho_b\sum_{\alpha\beta}
\left(\sigma_{\alpha\beta}^{ab}\right)^2Tr \left\{\bt_{\alpha\beta}^{ab}
\balpha \left[\partial\bg_{\alpha\beta}^{ab}-
\partial\bg_{\alpha\beta}^{(0)ab}\right]\balpha\right\},
\label{aaaa}
\ee
\be
\beta\left(P-P^{(ref)}\right)=-\beta{A-A^{(ref)}\over V}-{1\over 2}\sum_{ab}
\rho_a\rho_b\sum_{\alpha\beta}\left\{\balpha\left[\chat-\chat^{(0)}\right]
\balpha\right\}_{00}+\beta{\Delta\over V},
\label{pppp}
\ee
\be
-\beta\rho_a\left(\mu_\alpha^a-\mu_\alpha^{(ref)a}\right)=
{1\over 
2}\rho_a\sum_b\rho_b\sum_\beta\left\{\balpha\left[\chat_{\alpha\beta}^{ab}-
\chat_{\alpha\beta}^{(0)ab}\right]\balpha\right\}_{00}-
\beta{\Delta E_\alpha^a\over V}
\label{mumumu}
\ee
where the quantities with the superscript $(ref)$ denote the reference 
system quantities and the quantities with superscript $(0)$ denote the 
corresponding zero charge PMSA quantities
$$
\chat=\chat(k=0),\;\;\;
\chat_{\alpha\beta}^{ab}=\chat_{\alpha\beta}^{ab}(k=0),\;\;\;
\bg_{\alpha\beta}^{ab}=\bg_{\alpha\beta}^{ab}(\sigma_{\alpha\beta}^{ab}+),
\;\;\;
\partial \bg_{\alpha\beta}^{ab}={\partial \bg_{\alpha\beta}^{ab}
(r=\sigma_{\alpha\beta}^{ab}+)\over \partial r},
$$
\be
{\Delta E_\alpha^a\over V}=2\pi\rho_a\sum_b\rho_b\sum_\beta\int_0^\infty
r^2g_{\alpha\beta}^{ab}(r) U_{\alpha\beta}^{(C)ab}(r)\;dr
\label{egen}
\ee
and
\be
J=-{1\over 6}\beta\sum_{ab}\rho_a\rho_b\sum_{\alpha\beta}
\int_{r>\sigma_{\alpha\beta}^{ab}+} g_{\alpha\beta}^{ab}(r){\bf r}
\nablab U_{\alpha\beta}^{(C)ab}(r)\;d{\bf r}
\label{jjjj}
\ee
The reference system is represented by the multicomponent mixture of 
uncharged hard-sphere chain molecules. Expressions (\ref{aaaa}), 
(\ref{pppp}) and (\ref{mumumu}) are quite general and apply for any type of 
the potential outside the hard-core. The quantities, which enter these 
expressions can be written in terms of the present solution of PMSA for the 
model at hand. We have
\be
\beta{\Delta E_\alpha^a\over V}=\beta^*\rho_az_\alpha^aN_\alpha^a,
\hspace{8mm}
J={1\over 3}{\Delta E\over V},
\label{td1}
\ee
$$
\sigma_{\alpha\beta}^{ab}\partial \bg_{\alpha\beta}^{ab}=
-{\bf h}_{\alpha\beta}^{ab}+
\sum_d\rho_d\sum_\gamma\sigma_{\alpha\gamma}^{ad}{\bf h}_{\alpha\gamma}^{ad}
\balpha\left[\bt_{\gamma\beta}^{db}-
(\bzh_\gamma^d)^T\bah_\beta^b\right]-
\bE\balpha\sum_d\rho_d
\sum_\alpha\lambda_{\gamma\alpha}^{da}\bt_{\gamma\beta}^{db}-
$$
\be
-{1\over 2}\sum_d
\rho_d\sum_\gamma\sigma_\gamma^d\left\{\left[
{1\over \pi}\bt_{\alpha\gamma}^{ad}\balpha-{1\over 6}(\sigma_\gamma^d)^2
{\bf 1}\right]\bEh^T\bAh_\beta^b+
\left[2\sigma_{\alpha\gamma}^{ad}\bg_{\alpha\gamma}^{ad}-\bE\sigma_\gamma^d-
{1\over \pi\sigma_\gamma^d}\bt_{\alpha\gamma}^{ad}\right]\balpha
\bbeta_{\gamma\beta}^{db}\right\},
\label{td2}
\ee
\be
\sum_{ab}\rho_a\rho_b\sum_{\alpha\beta}\left[\balpha\chat_{\alpha\beta}^{ab}
\balpha\right]_{00}=
\sum_{ab}\rho_a\rho_b\sum_{\alpha\beta}\left\{\balpha\left[
(\bM_{\beta\alpha}^{ba})^T+\bM_{\alpha\beta}^{ab}+\sum_c\rho_c\sum_\gamma
\bM_{\alpha\gamma}^{ac}\balpha (\bM_{\beta\gamma}^{bc})^T\right]
\balpha\right\}_{00},
\label{td3}
\ee
where
\be
\bM_{\alpha\beta}^{ab}=-{1\over 12}(\sigma_\alpha^a)^3\bEh^T\bAh_\beta^b-
{1\over 2}(\sigma_\alpha^a)^2\bbeta_{\alpha\beta}^{ab}+\sigma_\alpha^a
\bt_{\alpha\beta}^{ab}+\lambda_{\beta\alpha}^{ba}(\bzh_\alpha^a)^T
\bah_\beta^b.
\label{td4}
\ee
Here expression (\ref{td3}) is obtained by differentiating of the equation 
(\ref{ozf4}) twice and considering the limit of 
$r\rightarrow \sigma_{\alpha\beta}^{ab}+$, and expression (\ref{td4}) 
follows from the first of the equations (\ref{ozk}) at $k\rightarrow 0$.

\section{Results and discussion}

To illustrate the above solution of the PMSA in this section we present the 
numerical results for thermodynamical properties of the several versions of 
charged chain fluid model. We consider two-component mixture of chain 
polyions with the beads of equal charge $z_\alpha^p=z_p=-1$ and oppositely 
charged counterions $z_\alpha^c=z_c=1$ (model M1) and two versions of 
three-component mixture of chain polyions with diblock and alternating 
distribution of oppositelly charged beads, $z_+^p=-z_-^p=1$, and two types 
of counterions with opposite charges $z_+^c=-z_-^c=1$ (models M2 and M3). 
The densities of the counterions of these two models were choosed to be 
$\rho^c_+=\rho^c_-={1\over 2}m_p\rho_p$. Here the indices $p$ and $c$ denote 
polyion and counterion, respectively. We also consider the one-component 
version of charged chain fluid models with diblock and alternating 
distribution of the charge along the chain backbone. In all cases the 
hard-sphere diameters of the chain beads $\sigma_\alpha^p$ and counterions 
$\sigma_\alpha^c$ were choosed to be equal, 
$\sigma_\alpha^p=\sigma_\alpha^c=\sigma=1$. Schematic representation of the 
models studied are shown in figure 1. Thermodynamical properties of the 
uncharged versions of these models were calculated using TPTD theory of 
Chang and Sandler \cite{chang2}.

In figure 2 we compare osmotic coefficient results obtained using the 
present version of the PMSA, von Solms and Chiew version of PMSA 
\cite{solms} and molecular dynamic (MD) computer simulation method 
\cite{kremer} for the model M1 at a different chain length ($m_p=16,32$ and 
$64$) and Bjerrum length $\lambda_B=\beta^*=0.833$. In the region of 
the intermediate densities our theory underestimate and the theory of von 
Solms and Chiew overestimate the value of the osmotic coefficient 
$\phi=\beta P/\rho$, where $\rho=\sum_a\rho_a$. In the region of the low 
densities both theories predict for the osmotic coefficient its ideal gas 
value of $\phi=0$, while MD simulation gives the value, which is slightly 
above 1. This disagreement could be due to the uncertainties of the MD 
simulation in the diluted region. In general predictions of the present 
approach are more accurate than those of von Solms and Chiew approach 
\cite{solms}.

In figures 3-5 we compare the density dependence of the osmotic pressure and 
osmotic coefficient at different values of Bjerrum length 
($\lambda_B=0.833,2.499$) for different polyelectrolyte models. In the case 
of low Bjerrum length ($\lambda_B=0.833$) one can see the linear dependence 
of the osmotic pressure with respect to the density on the $log-log$ scale 
(figures 3a-5a). In this region the osmotic pressure is almost independent 
of the chain length and distribution of the charge along the chain. With the 
increase of $\lambda_B$ ($\lambda_B=2.499$) there is a departure from the 
linearity, which is substantial for the model M1, slightly smaller for the 
model M2 and almost negligible for the model M3. At the same time one can 
see the substantial chain length dependence of the osmotic pressure in the 
case of the model M2. This dependence becomes smaller in the case of the 
model M1 and is absent for the model M3 (figures 3b-5b). Thus for the 
systems with more random distribution of the charge along the chain backbone 
the chain length dependence is smaller. With the increase of the chain 
length this dependence becomes weaker and it is reasonable to expect, that 
for $m_p>64$ it will be negligible for all the models studied.

In figur 6 the liquid-gas phase diagrams for the one-component 
version of the models with diblock and alternating distribution of the 
charge are presented. We consider the systems with molecular chain length 
$m_p=2,4,6,8,10$. Computer simulation results are available only in the case 
of the chains length $m_p=2$ \cite{patey1}. In general theoretical results 
are in qualitative agreement with computer simulation results with certain 
disagreement for the slope steepness of the liquid branches of the 
coexistence curves; the slope of theoretical curve is steeper than that of 
the computer simulation curve. In addition, the position of the theoretical 
critical point is shifted towards lower values of the density and higher 
values of the temperature. According to the previous discussion the systems 
with diblock distribution of the charge have larger degree of nonideality in 
comparison with the systems with alternating distribution of the charge. 
This is reflected in the figure 6, where the critical temperature of 
the systems with diblock chains is higher than that of the systems with 
alternating chains of the same length. The corresponding values of the 
critical densities is larger in the case of systems with alternating chains. 
In both cases increase of the chain length leads to the increase of the 
critical temperature and slight decrease (increase) of the critical density 
for diblock (alternating) distribution of the charge. We note in passing, 
that due to the lack of the solution of the equation for screening parameter 
$\Gamma$ (\ref{gamma}) at lower densities we were not able to determine the 
coexistance curve for the systems with alternating chain charge of the chain 
length $m_p=8,10$ below the temperatures shown in figure 6b.

Finally in figure 7 we present the liquid-gas phase diagram for the system 
of trimers with the middle bead charge $z_2^p=-2z_1^p=-2z_3^p$, where 
$z_1^p$ and $z_3^p$ are the charges of the terminal beads, $z_1^p=z_3^p=1$. 
Similar as in the case of alternating charged chain system with $m_p=8,9$ 
there is no solution of the equation (\ref{gamma}) in the low density 
region, therefore we were not able to calculate the coexistence curve for 
the temperatures lower than those shown in the figure. This model can be 
seen as a complete association limit of the primitive 2:1 electrolyte model. 
Recent computer simulation studies \cite{19,patey1,21,patey2} of the 
electrolyte primitive model suggest that in the vicinity of the coexistance 
region the system is highly associated and fraction of charged clusters, 
including free ions, is negligible. Therefore the equilibrium properties of 
the system are determined by the presence of the neutral ionic clusters, to 
which the lower order clusters give the main contribution. Assuming that in 
the case of the electrolyte restricted primitive model (RPM) all such 
clusters are represented by the neutral ionic pairs the RPM liquid-gas phase 
diagram has been successfully reproduced by the Monte Carlo (MC) computer 
simulation method \cite{patey1} and theoretically \cite{ktd}. Making similar 
assumption in the case of the 2:1 primitive electrolyte model the 
corresponding liquid-gas phase diagram can be modeled using the phase 
diagram obtained for trimers. Unfortunately computer simulation predictions 
for the entire phase diagram of the 2:1 electrolyte model currently is not 
available. Recently Camp and Patey \cite{patey2} present MC estimate for the 
critical temperature, which appeares to be $T^*_c=0.1\pm 0.01$. The rough 
estimate of the ranges for the critical density can be obtained from the 
figure 5 of \cite{patey2}; our estimate is $\rho_c^*=0.115\pm 0.085$. This 
estimate for the critical point together with Debye-H\"uckel (DH) 
theory estimate \cite{patey2} and MSA phase diagram are shown in figure 7. 
In addition, for the sake of completeness, we present the liquid-gas phase 
diagram for the RPM of electrolyte, obtained from the MC simulation method 
\cite{19,32}, MSA and our PMSA for neutral diatomics \cite{ktd}. One can see 
that predictions of the PMSA is much better in comparison with usual MSA. In 
both cases DH predictions for the critical temperature is close to those of 
the PMSA, however the corresponding predictions for the critical density are 
much worse.

\section{Concluding remarks}

In this study we obtain an analytical solution of the ideal chain PMSA for 
the multicomponent mixture of charged hard-sphere flexible linear chain 
molecules. This solution apply to any mixture of chain molecules with 
arbitrary distribution of the charge and size of the beads along the 
molecular backbone. We present closed form analytical expressions for 
thermodynamical and structure properties of the system. These expressions
are used to calculate thermodynamics of several different versions of 
charged hard-sphere chain model. We consider two-component mixture of chains 
with equally charged beads and oppositely charged counterions, two versions 
of the three-component mixture of chains with diblock and alternating 
distribution of the charge and two types of oppositely charged counterions. 
Comparison of the theoretical predictions for the osmotic pressure of the 
two-component model with corresponding computer simulation predictions shows 
that the present version of the ideal chain PMSA theory is reasonably 
accurate and gives somewhat better agreement than that of the theory of 
von Solms and Chiew \cite{solms}. Effects due to the difference in Bjerrum 
length, molecular chain length and distribution of the charge was also 
studied. The largest deviation from ideal behaviour was demonstrated by the 
system with uniform distribution of the charge; behaviour of the system with 
alternating distribution of the charge shows much less nonideality. This 
conclusion is varyfied by the comparison of the liquid-gas phase diagrams 
calculated for the one-component version of the models with diblock and 
alternating distribution of the charge; the latter model has substatially 
lower critical temperature than that of the former. With the increase of the 
chain length the critical temperature for both models increases, while the 
corresponding critical density slightly increases for alternating charge 
model and decreases for diblock charge model. Finally it was demonstrated, 
that liquid-gas phase diagram of the neutral trimers with doubly charged 
middle bead and singly charged terminal beads of the opposite sighn can be 
used to correct MSA results for the phase diagram of 2:1 primitive 
electrolyte model.\\
\\
{\centerline {\bf ACKNOWLEDGMENTS}}

\pagebreak

{\bf CAPTIONS TO THE FIGURES}
\vspace{10mm}

{\bf Figure 1.}  Schematic diagrams of the models with uniform 
distribution of the charge (M1), diblock distribution of the charge (M2) and 
alternating distribution of the charge.

{\bf Figure 2.}  Osmotic coefficient $\phi=PV/NkT$ as a function of the 
packing fraction $\eta=\pi\rho\sigma^3/6$ for the model M1 at 
$\lambda_B=0.833$ for $m_p=16$ (a), $m_p=32$ (b) and $m_p=64$ (c). 
Predictions of the present theory (solid lines), von Solms and Chiew theory 
\cite{solms} (dashed lines) and computer simulation predictions 
\cite{kremer} (diamons).

{\bf Figure 3.}  Osmotic pressure $P^*=\beta P\sigma^3$ (a) and osmotic 
coefficient  $\phi=PV/NkT$ (b) as a function of packing fraction 
$\eta=\pi\rho_T\sigma^3/6$ ($\rho_T=\sum_am_a\rho_a$ for the model M1 from 
the present theory at $\lambda_B=2.499$ (solid lines), $\lambda_B=0.833$ 
(dashed lines) and  $\lambda_B=0$ (dashed-dotted lines). For each set of 
lines from the top to the bottom at $\log{\eta}=-2$ $m_p=8,16,32,64$. 
Symbols are computer simulation predictions \cite{kremer} for 
$\lambda_B=0.833$ and for $m_p=16$ (diamonds), $m_p=32$ (circles) and 
$m_p=64$ (squares).

{\bf Figure 4.} The same as in figure 3 for the model M2.

{\bf Figure 5.} The same as in figure 3 for the model M3.

{\bf Figure 6.} Liquid-gas phase diagram in $T^*=1/\beta^*$ versus 
reduced density of the beads $\rho^*_b=m_p\rho\sigma^3$ coordinates for one 
component chain fluids with diblock distribution of the charge (a) and with 
alternating distribution of the charge (b). From the top to the bottom at 
$\rho^*_b=0.05$ $m_p=10,8,6,4,2$. Solid lines are predictions from the 
present theory and diamonds are computer simulation predictions 
\cite{patey1} for $m_p=2$.

{\bf Figure 7} Liquid-gas phase diagram in $T^*=1/\beta^*$ versus reduced 
ionic density $\rho^*=(\rho_++\rho_-)\sigma^3$ coordinates for restricted 
primitive model (RPM) of electrolyte (lower portions of the figure for 
$T^*>0.08$) and for the primitive model (PM) of electrolyte with 2:1 
assymetry in charge (upper portions of the figure for $T^*<0.08$). 
Present theory (solid lines), MSA (dashed lines), Debye-H\"uckel theory 
\cite{patey2} (squares), computer simulation estimate for the critical point 
of 1:2 electrolyte PM \cite{patey2} (open diamonds) and for the phase 
diagram of electrolyte RPM (open circles \cite{19}, solid diamonds 
\cite{32}).

\clearpage

\begin{figure}
\end{figure}

\unitlength=1mm
\centerline{
\begin{picture}(95,80)
\put(5,65){\makebox(0,0){M1}}
\put(15,65){\makebox(0,0){$\cdots$}}
\put(23,65){\circle{10}}
\put(23,65){\makebox(0,0){$-$}}
\put(30.072,72.072){\circle{10}}
\put(30.072,72.072){\makebox(0,0){$-$}}
\put(37.144,65){\circle{10}}
\put(37.144,65){\makebox(0,0){$-$}}
\put(47.144,65){\circle{10}}
\put(47.144,65){\makebox(0,0){$-$}}
\put(56.5,65){\makebox(0,0){$\cdots\;$,}}
\put(78,65){\circle{10}}
\put(78,65){\makebox(0,0){$+$}}
\put(86,65){\makebox(0,0){$;$}}
\put(5,40){\makebox(0,0){M2}}
\put(15,40){\makebox(0,0){$\cdots$}}
\put(23,40){\circle{10}}
\put(23,40){\makebox(0,0){$-$}}
\put(30.072,47.072){\circle{10}}
\put(30.072,47.072){\makebox(0,0){$-$}}
\put(37.144,40){\circle{10}}
\put(37.144,40){\makebox(0,0){$+$}}
\put(47.144,40){\circle{10}}
\put(47.144,40){\makebox(0,0){$+$}}
\put(56.5,40){\makebox(0,0){$\cdots\;$,}}
\put(70,40){\circle{10}}
\put(70,40){\makebox(0,0){$+$}}
\put(77,40){\makebox(0,0){$,$}}
\put(85,40){\circle{10}}
\put(85,40){\makebox(0,0){$-$}}
\put(92,40){\makebox(0,0){$;$}}
\put(5,15){\makebox(0,0){M3}}
\put(15,15){\makebox(0,0){$\cdots$}}
\put(23,15){\circle{10}}
\put(23,15){\makebox(0,0){$-$}}
\put(30.072,22.072){\circle{10}}
\put(30.072,22.072){\makebox(0,0){$+$}}
\put(37.144,15){\circle{10}}
\put(37.144,15){\makebox(0,0){$-$}}
\put(47.144,15){\circle{10}}
\put(47.144,15){\makebox(0,0){$+$}}
\put(56.5,15){\makebox(0,0){$\cdots\;$,}}
\put(70,15){\circle{10}}
\put(70,15){\makebox(0,0){$+$}}
\put(77,15){\makebox(0,0){$,$}}
\put(85,15){\circle{10}}
\put(85,15){\makebox(0,0){$-$}}
\put(92,15){\makebox(0,0){$.$}}
\put(85,-105){\makebox(0,0){Figure 1 (Kalyuzhnyi and Cummings)}}
\end{picture}}

\clearpage

\begin{figure}
\epsfxsize 120mm
\epsfysize 67mm
\centerline{\epsffile{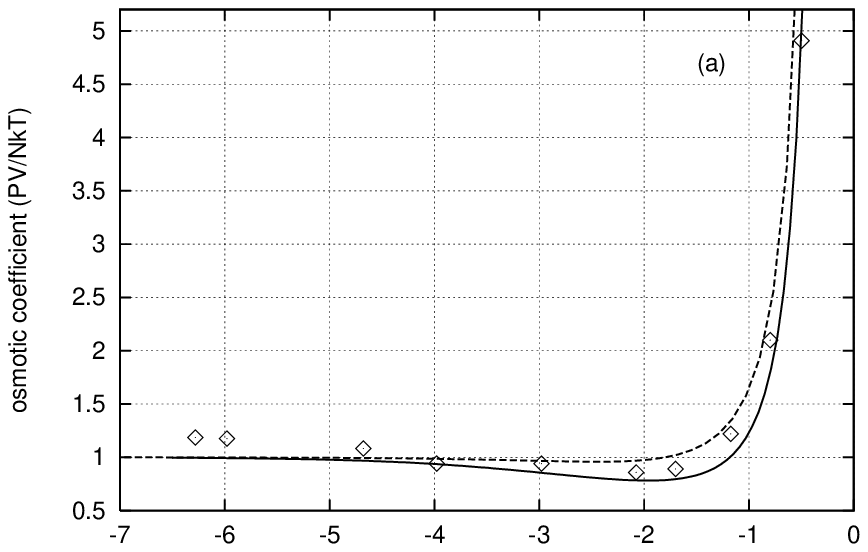}}
\end{figure}

\begin{figure}
\epsfxsize 120mm
\epsfysize 67mm
\centerline{\epsffile{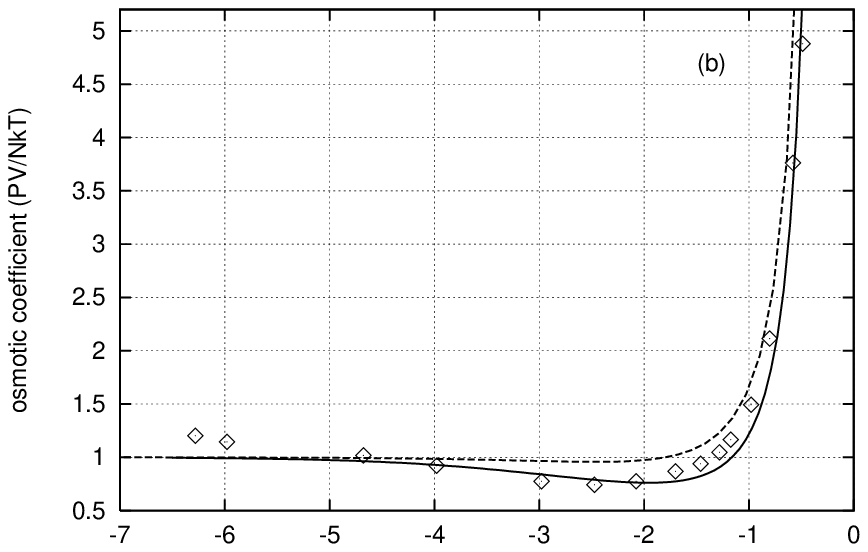}}
\end{figure}

\begin{figure}
\epsfxsize 120mm
\epsfysize 67mm
\centerline{\epsffile{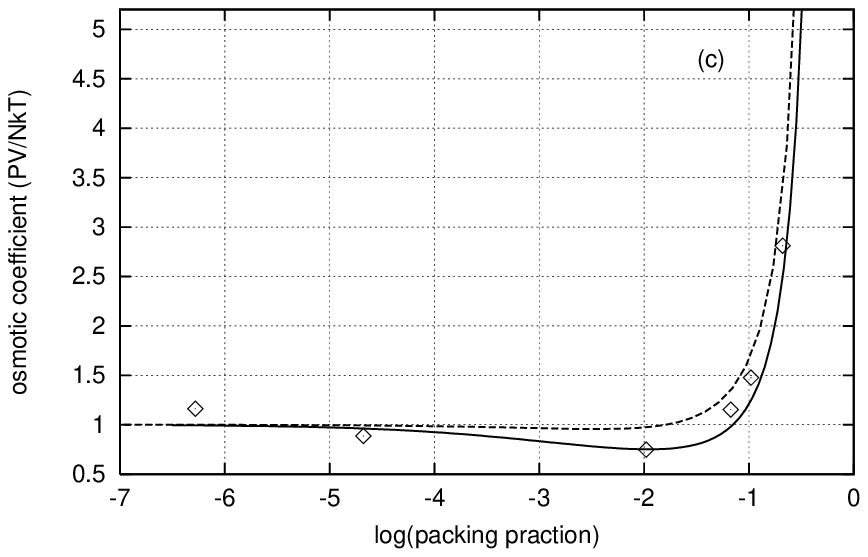}}
\vspace{2mm}
\hspace{85mm}
Figure 2 (Kalyuzhnyi and Cummings)
\end{figure}

\clearpage

\begin{figure}
\epsfxsize 120mm
\epsfysize 75mm
\centerline{\epsffile{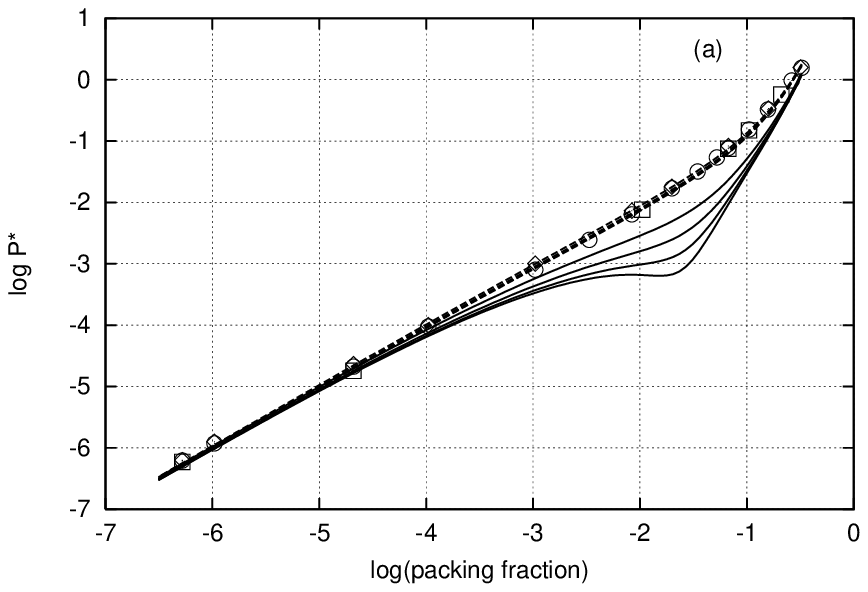}}
\end{figure}

\begin{figure}
\epsfxsize 120mm
\epsfysize 75mm
\centerline{\epsffile{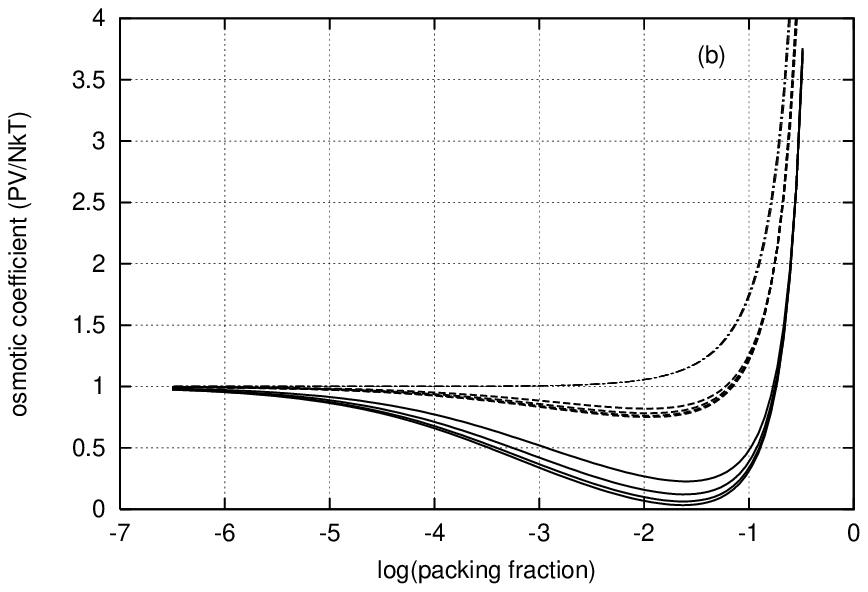}}
\vspace{15mm}
\hspace{85mm}
Figure 3 (Kalyuzhnyi and Cummings)
\end{figure}

\clearpage

\begin{figure}
\epsfxsize 120mm
\epsfysize 75mm
\centerline{\epsffile{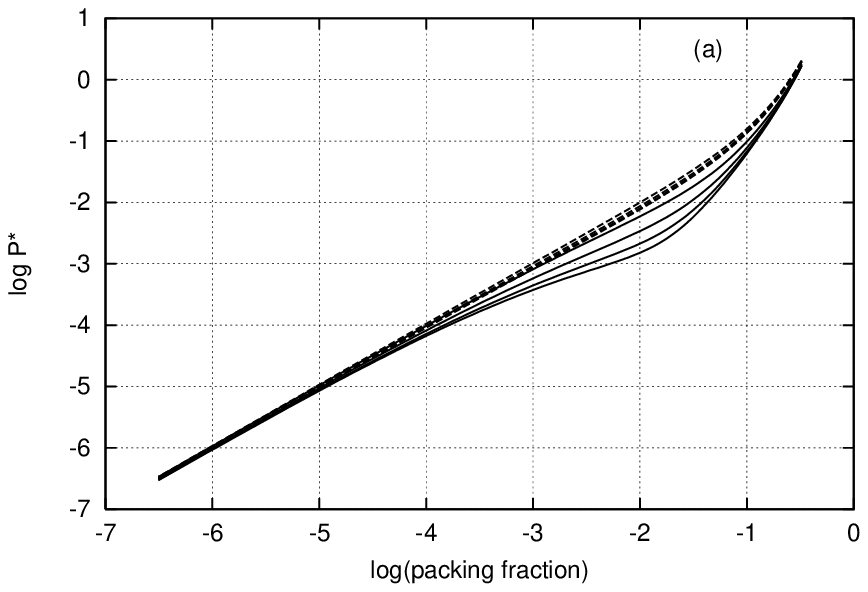}}
\end{figure}

\begin{figure}
\epsfxsize 120mm
\epsfysize 75mm
\centerline{\epsffile{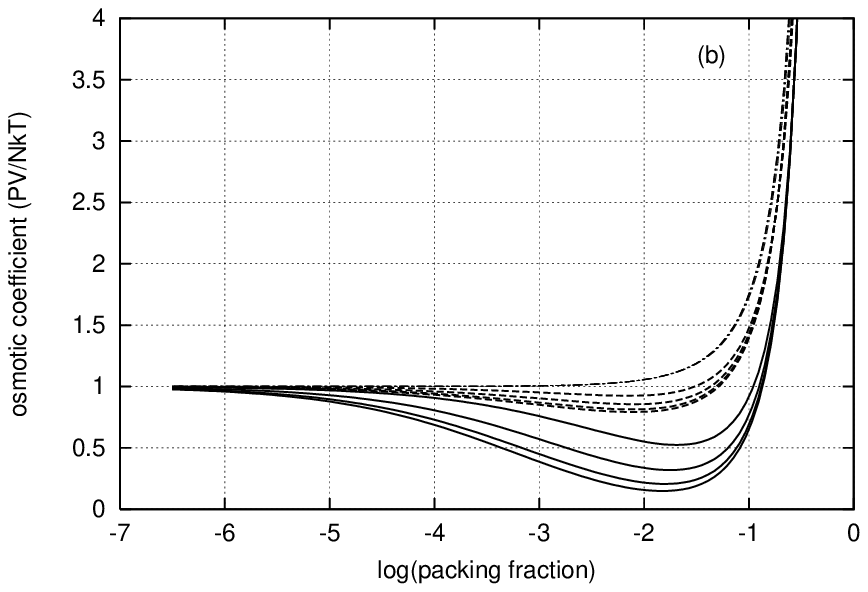}}
\vspace{15mm}
\hspace{85mm}
Figure 4 (Kalyuzhnyi and Cummings)
\end{figure}

\clearpage

\begin{figure}
\epsfxsize 120mm
\epsfysize 75mm
\centerline{\epsffile{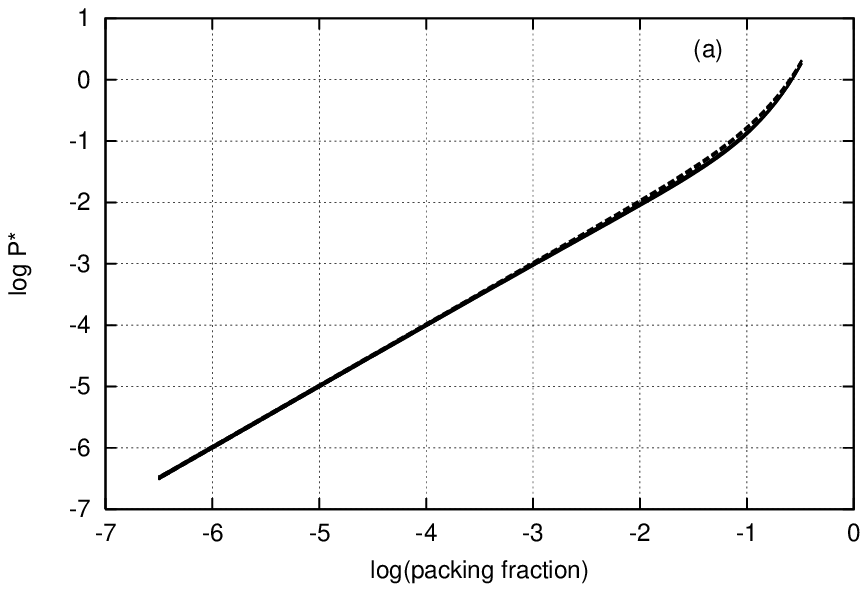}}
\end{figure}

\begin{figure}
\epsfxsize 120mm
\epsfysize 75mm
\centerline{\epsffile{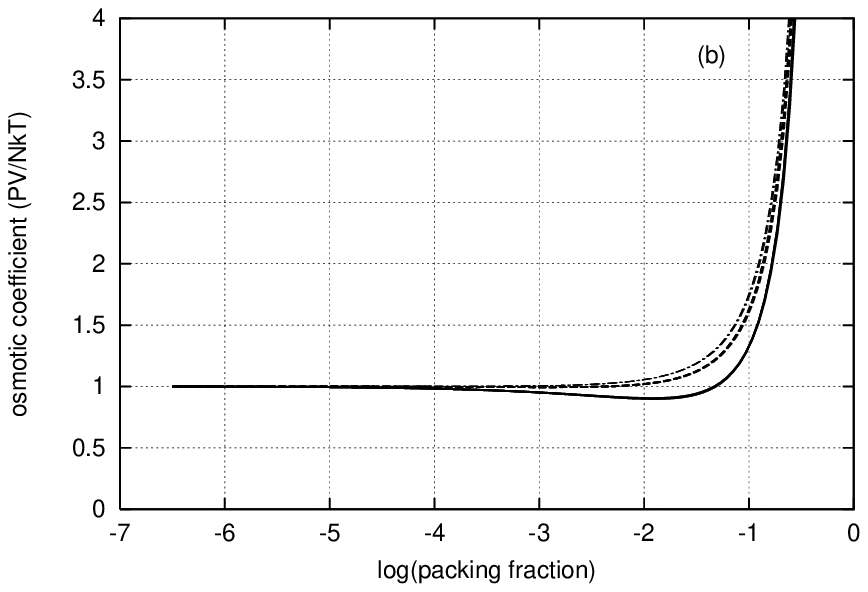}}
\vspace{15mm}
\hspace{85mm}
Figure 5 (Kalyuzhnyi and Cummings)
\end{figure}

\clearpage

\begin{figure}
\epsfxsize 120mm
\epsfysize 75mm
\centerline{\epsffile{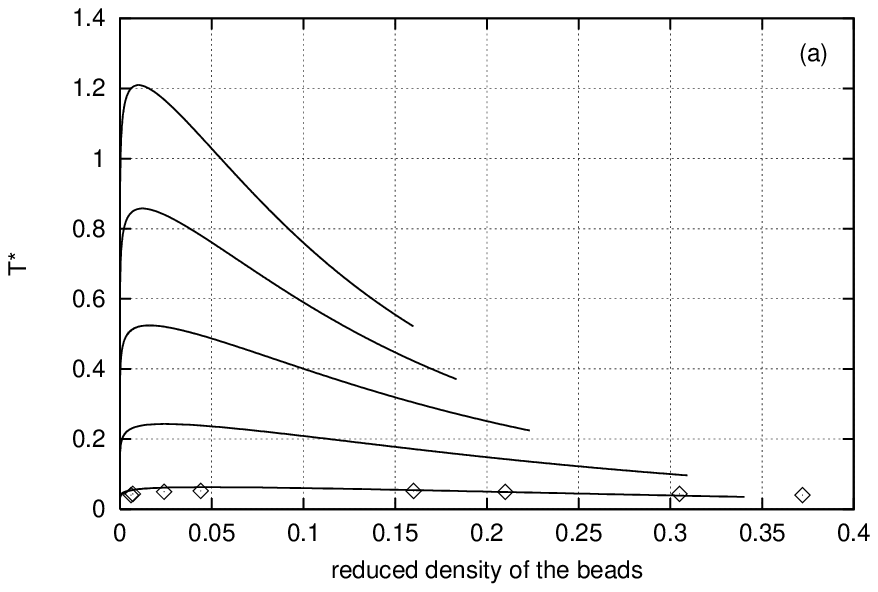}}
\end{figure}

\begin{figure}
\epsfxsize 120mm
\epsfysize 75mm
\centerline{\epsffile{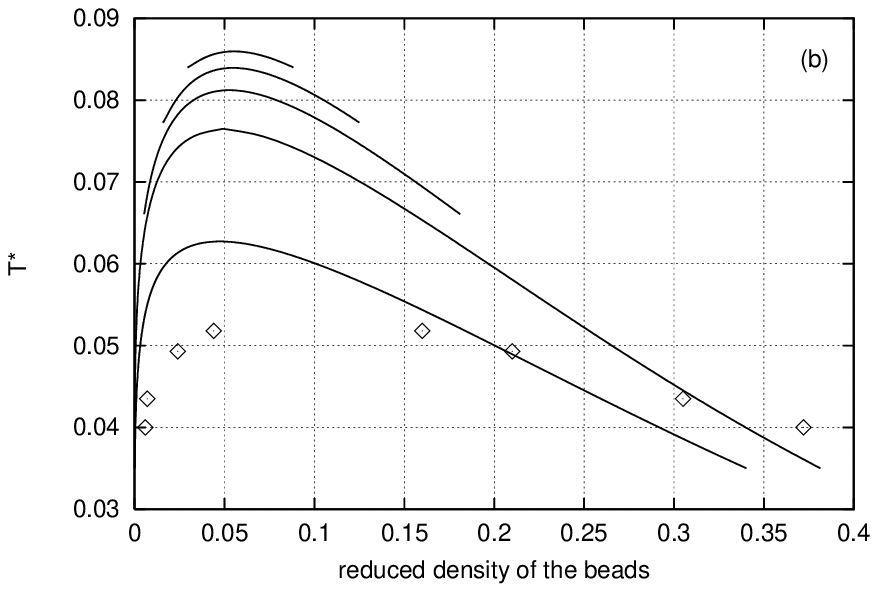}}
\vspace{15mm}
\hspace{85mm}
Figure 6 (Kalyuzhnyi and Cummings)
\end{figure}

\clearpage

\begin{figure}
\epsfxsize 120mm
\epsfysize 75mm
\centerline{\epsffile{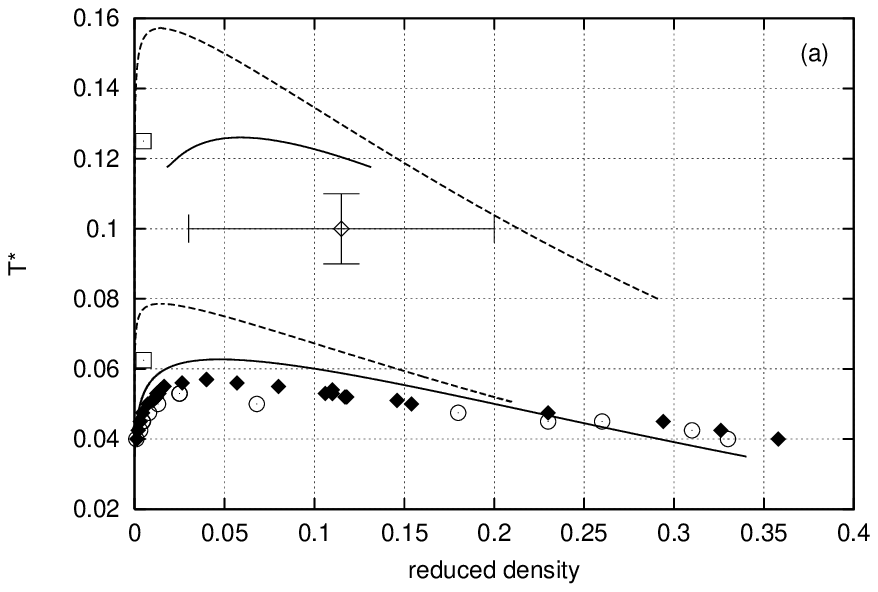}}
\end{figure}

\begin{figure}
\epsfxsize 120mm
\epsfysize 75mm
\centerline{\epsffile{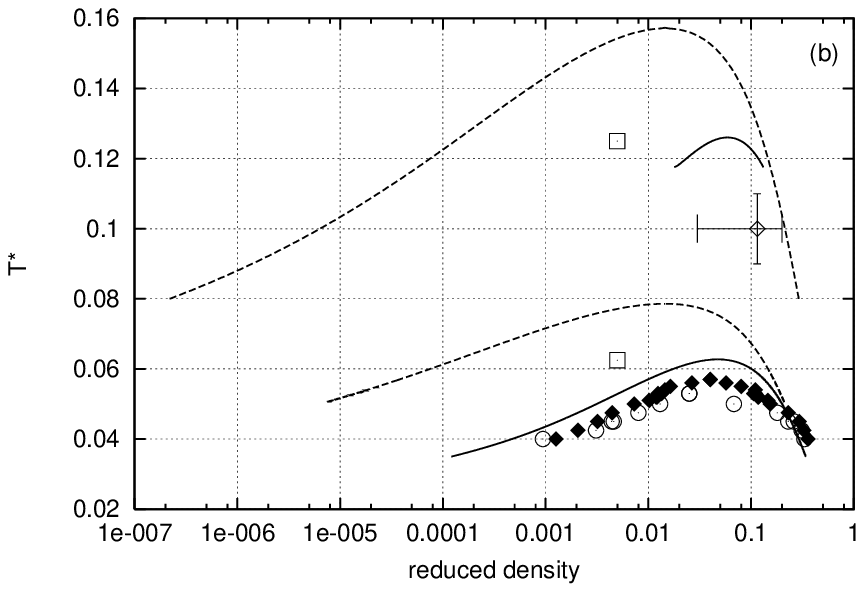}}
\vspace{15mm}
\hspace{85mm}
Figure 7 (Kalyuzhnyi and Cummings)
\end{figure}


\begin{references}

\bibitem{belloni} M. Dymitrowska, and L. Belloni, J. Chem. Phys. 
   {\bf 111}, 6633(1999).

\bibitem{currev} K. S. Schweizer, and J. G. Curro, Adv. Chem. Phys. 
   {\bf 98}, 1(1997).

\bibitem{china1} J. W. Jiang, Y. L. Liu, Y. Hu, and J. M. Prausnitz,
   J. Chem. Phys. {\bf 108}, 780(1998).

\bibitem{china2} J. Jiang, H. Liu, and Y. Hu, J. Chem. Phys.
   {\bf 110}, 4952(1999).

\bibitem{stellzhou} G. Stell, and Y. Zhou, J. Chem. Phys.
  {\bf 91}, 3618(1989); Y. Zhou, and G. Stell, {\it ibid} {\bf 96}, 
  1504(1992); {\bf 96}, 1507(1992); {\bf 102}, 8089(1995).

\bibitem{kstellion} Yu. V. Kalyuzhnyi, and G. Stell, Chem. Phys. Lett.
  {\bf 240}, 157(1995).

\bibitem{kblum0} L. Blum, Yu. V. Kalyuzhnyi, O. Bernard, and 
  J. N. Herrera-Pacheco, J. Phys.: Cond. Matter {\bf 8}, A143(1996).

\bibitem{kblum} I. A. Protsykevytch, Yu. V. Kalyuzhnyi, M. F. Holovko, 
   and L. Blum, J. Mol. Liquids {\bf 73,74}, 1(1997).

\bibitem{ktd} Yu. V. Kalyuzhnyi, Mol. Phys. {\bf 94}, 735(1998).

\bibitem{solms} N. von Solms, and Y. C. Chiew, J. Chem. Phys.
   {\bf 111}, 4839(1999).

\bibitem{w34} M. S. Wertheim, J. Stat. Phys. {\bf 42}, 459; {\it ibid}
   {\bf 42}, 477(1986).

\bibitem{holkal} M. F. Holovko, and Yu. V. Kalyuzhnyi, Mol. Phys.
   {\bf 73}, 1145(1991).

\bibitem{chang1} J. Chang, and S. I. Sandler, J. Chem. Phys. {\bf 102},
   437(1995).

\bibitem{chang2} J. Chang, and S. I. Sandler, J. Chem. Phys.
  {\bf 103}, 3196(1995).  

\bibitem{kalcum} Yu. V. Kalyuzhnyi, and P. T. Cummings, J. Chem. Phys.
   {\bf 104}, 3325(1996).

\bibitem{rossky} P. J. Rossky, and R. A. Chiles, Mol. Phys. {\bf 51}, 661(1984).

\bibitem{csl} D. Chandler, R. Silbey, and Ladanyi, Mol. Phys.
  {\bf 46} 1335(1982).

\bibitem{kstellsym} Yu. V. Kalyuzhnyi, and G. Stell, Mol. Phys.
  {\bf 78}, 1247(1993).

\bibitem{stellphys} G. Stell, Physica a {\bf 231}, 1(1996).

\bibitem{kalcumon} Yu. V. Kalyuzhnyi, and P. T. Cummings, J. Chem. Phys.
  {\bf 105}, 2011(1996).

\bibitem{kallin1} Yu. V. Kalyuzhnyi, C.-T. Lin, and G. Stell, J. Chem. Phys.
  {\bf 108}, 6525(1998).

\bibitem{hoyestell} J. S. H\o ye and G. Stell, J.Chem.Phys. {\bf 67}, 
  439(1977).

\bibitem{kremer} M. J. Stevens, and K. Kremer, J. Chem. Phys. 
  {\bf 103}, 1669(1995).

\bibitem{berblumpes} O. Bernard, and L. Blum, J. Chem. Phys.
  {\bf 112}, 7227(2000).

\bibitem{blumrpm} L. Blum, Mol. Phys. {\bf 30}, 1529(1975); L. Blum, and 
  J. S. H\o ye, J. Phys. Chem. {\bf 81}, 1311(1977).

\bibitem{patey1} J. C. Shelley, and G. N. Patey, J. Chem. Phys. 
  {\bf 103}, 8299(1995).

\bibitem{19} A. Z. Panagiotopoulos, Fluid Phase Equil. {\bf 76}, 97(1992).

\bibitem{21} J.-M., Caillol, and J.-J. Weiss, J. Chem. Phys. 
  {\bf 102}, 7610(1995).

\bibitem{patey2} P. J. Camp, and G. N. Patey, J. Chem. Phys. 
  {\bf 111}, 9000(1999).

\bibitem{32} J. M. Caillol, J. Chem. Phys. {\bf 100}, 2161(1994).

\end{references}
\end{document}